# Estimation de l'indice des vides minimum des mélanges binaires de sols : approche théorique

*Estimation of the minimum void ratio of binary soil mixtures : theoretical approach*

Gérard ROQUIER[1]
[1] *Université de Poitiers (ENSI Poitiers), CNRS, IC2MP, HydrASA, Poitiers, France*

**RÉSUMÉ –** L'indice des vides minimum des mélanges sable-fines en fonction de la teneur en fines est une propriété importante à connaître en ingénierie géotechnique. Dans cet article, un modèle mathématique est proposé pour l'estimer. Il tient compte à la fois de l'insertion des fines dans le squelette granulaire du sable lorsque le rapport de tailles le permet et du mode d'assemblage des particules.

**ABSTRACT –** The minimum void ratio of sand-fine mixtures as a function of fines content is an important property to know in geotechnical engineering. In this paper, a mathematical model is proposed to estimate it. It takes into account both the insertion of fines into the granular skeleton of the sand when the size ratio allows it, and the way in which the particles are assembled.

## 1. Introduction

A l'heure où les phénomènes climatiques sont de plus en plus violents et soudains, la compréhension des écoulements granulaires (glissement de terrain, éboulements) constitue un enjeu majeur en géotechnique. Pour prédire la propagation d'un écoulement, les travaux se consacrent notamment aux écoulements denses lors de la transition de blocage séparant un état fluide d'un état solide. La compacité de blocage est gouvernée par la configuration de l'empilement des grains, qui dépend de nombreux facteurs : forme des particules, friction/cohésion entre particules, polydispersité, l'état de contraintes, etc.
Un nombre considérable de publications concerne la modélisation de la compacité d'empilements binaires, permettant d'étudier aussi bien le potentiel de liquéfaction d'un sable limoneux que sa perméabilité. La plupart d'entre elles sont largement basées sur des approches phénoménologiques mais relativement peu adoptent une approche plus fondamentale. Le Modèle Théorique de Compacité (Theoretical Packing Density Model, TPDM), au contraire, repose sur un corpus faisant appel à des paramètres ayant une signification physique. Validé pour les mélanges de sphères et de micro-poudres, les arrangements binaires cristallins et les distributions granulaires liées à la formulation des bétons et des asphaltes, le TPDM est testé dans cet article pour estimer l'indice des vides minimum de mélanges binaires sable-fines non plastiques, omniprésents dans la nature et dans certaines réalisations humaines (digues, remblais). Dans ce qui suit, les gros grains et les petits grains désigneront respectivement le sable et les fines non plastiques.

## 2. Le concept du Modèle Théorique de Compacité (TPDM) pour les mélanges binaires

La compacité d'un mélange granulaire représente le rapport entre le volume solide et le volume total d'un empilement alors que l'indice des vides correspond au rapport entre le volume des vides et le volume solide. Le but du TPDM est d'estimer la compacité ou l'indice des vides d'un mélange de populations de grains de taille identique (les classes granulaires) à partir de la connaissance de leurs proportions mutuelles, de la compacité ou de l'indice des vides de chacune des classes empilées isolément et d'une grandeur caractérisant le processus de création de l'empilement.

## 2.1. Compacité virtuelle ou indice des vides virtuel pour un empilement binaire

Une distinction est tout d'abord établie entre compacité *virtuelle*, valeur maximale atteinte en construisant l'édifice grain par grain, et compacité *réelle*, obtenue dans une expérience d'empilement aléatoire. Par exemple, pour des sphères parfaites de même taille, la compacité *virtuelle* vaut $\pi/(3\sqrt{2}) \approx 0.74$ alors que la compacité *réelle* varie entre 0.60 et 0.64 suivant l'énergie mise en œuvre.

La compacité *virtuelle* d'un mélange binaire « sans interaction » (dont le contraste des tailles est très important) est tout d'abord calculée en distinguant le cas des gros dominants de celui des petits dominants. Dans le premier cas, le volume des gros grains, rapporté à un volume total unité, est égal à leur compacité lorsqu'ils sont empilés isolément. Dans le deuxième cas, les petits grains remplissent un récipient constitué du volume total, diminué de celui des gros grains. Dans la réalité, la dimension des petits grains n'est pas toujours très petite devant celle des gros grains : des interactions géométriques doivent être prises en compte. Celles-ci font l'objet d'un traitement théorique dans le TPDM : l'effet de *paroi* et l'effet de *desserrement* sont respectivement matérialisés par un coefficient $b_{TPDM}$ lorsque les petits grains sont dominants et par un coefficient $a_{TPDM}$ lorsque les gros grains sont dominants. Les deux s'expriment en fonction de $x$ qui représente le rapport des diamètres entre petits grains et gros grains. L'effet de *paroi* représente le surplus de porosité qui se manifeste lorsque des petites particules viennent se plaquer contre la paroi d'une particule plus grosse. L'effet de *desserrement* se produit lorsqu'une petite particule n'est pas suffisamment fine pour s'introduire entre des gros grains sans les desserrer. En admettant l'additivité de ces phénomènes et en supposant que les interactions granulaires dépendent principalement du rapport des tailles de grains, il est possible d'écrire le modèle mathématique correspondant à la compacité *virtuelle* $\gamma_i$ ou à l'indice des vides *virtuel* $e_i^*$ d'un mélange binaire, respectivement dans le cas des gros grains dominants ($\gamma_1$, $e_1^*$) et dans le cas des grains fins dominants ($\gamma_2$, $e_2^*$) :

$$\begin{cases} \gamma_1 = \dfrac{\beta_1}{1 - \left(1 - \dfrac{\beta_1}{\beta_2} a_{TPDM}\right) y_2} \\ \gamma_2 = \dfrac{\beta_2}{1 - \left(1 - \beta_2 + b_{TPDM}\, \beta_2 \left(1 - \dfrac{1}{\beta_1}\right)\right) y_1} \end{cases} \quad (1)$$

$$\begin{cases} e_1^* = e_1^v - \left(1 + e_1^v - a_{TPDM}(1 + e_2^v)\right) y_2 \\ e_2^* = e_2^v - (e_2^v - b_{TPDM}\, e_1^v) y_1 \end{cases} \quad (2)$$

où $\beta_1$ et $\beta_2$ sont les compacités *virtuelles* des classes granulaires grossière 1 et fine 2, $y_1$ et $y_2$ leurs proportions volumiques respectives, $b_{TPDM}$ et $a_{TPDM}$ les coefficients d'effet de *paroi* et d'effet de *desserrement*, $\gamma_1$ et $\gamma_2$ les compacités *virtuelles* du mélange binaire dans le cas des gros grains dominants et des grains fins dominants, $e_1^v$ et $e_2^v$ les indices des vides *virtuels* des classes granulaires grossière 1 et fine 2, $e_1^*$ et $e_2^*$ les indices des vides *virtuels* du mélange binaire dans le cas des gros grains dominants et des grains fins dominants.

Sans interactions géométriques (i.e. $a_{TPDM} = b_{TPDM} = 0$), les équations (2) s'écrivent :

$$\begin{cases} e_1^* = e_1^v - (1 + e_1^v)\, y_2 \\ e_2^* = e_2^v\, y_2 \end{cases} \quad (3)$$

Dans le premier scénario, les petites particules s'insèrent parfaitement dans les cavités existant dans le squelette des grosses particules : il s'agit du *mécanisme d'insertion*. Dans le second scénario, les grosses particules sont dispersées dans un océan de petites : il s'agit du *mécanisme de substitution*.

Dans le cas d'interaction totale (i.e. $a_{TPDM} = b_{TPDM} = 1$), les équations (2) s'écrivent :

$$\begin{cases} e_1^* = e_1^v\, y_1 + e_2^v\, y_2 \\ e_2^* = e_1^v\, y_1 + e_2^v\, y_2 \end{cases} \tag{4}$$

Les deux équations sont identiques : elles expriment la limite supérieure de l'indice des vides d'un mélange binaire lorsque les deux classes granulaires ont le même diamètre.

## 2.2. Compacité réelle ou indice des vides réel pour un empilement binaire

Le calcul de la compacité *virtuelle* ne constitue qu'une étape avant d'atteindre la compacité *réelle*, obtenue par un mode de mise en place donné. C'est pourquoi est définie une grandeur scalaire appelée indice de compaction $K$, représentative du processus de création de l'empilement. Elle se calcule à partir des proportions volumiques de chaque classe granulaire et des paramètres précédents déjà définis :

$$\begin{cases} K = K_1 + K_2 = \dfrac{\frac{y_1}{\beta_1}}{\frac{1}{\phi} - \frac{1}{\gamma_1}} + \dfrac{\frac{y_2}{\beta_2}}{\frac{1}{\phi} - \frac{1}{\gamma_2}} \\ K = K_1 + K_2 = \dfrac{(1 + e_1^v)\, y_1}{e - e_1^*} + \dfrac{(1 + e_2^v)\, y_2}{e - e_2^*} \end{cases} \tag{5}$$

où $K_1$ et $K_2$ sont les indices de compaction partiels des classes 1 et 2, $\phi$ et $e$ la compacité et l'indice des vides du mélange binaire recherchés. Si le processus d'empilement est parfait, $K$ tend vers l'infini et l'indice des vides du mélange binaire est bi-linéaire en fonction de la proportion volumique des petits grains : le modèle est alors dit linéaire. Dans le cas contraire, $K$ évolue dans une gamme généralement comprise entre 3-4 pour un simple déversement et 100 pour un empilement cristallin : le modèle devient alors non-linéaire. Si la valeur de $K$ correspondant à un mode de mise en place est connue, l'expression (5) apparaît comme une équation implicite en $\phi$ ou $e$, qui admet une et une seule racine positive, et qui peut être résolument facilement numériquement.

## 2.3. Détermination théorique du coefficient d'effet de paroi et du coefficient d'effet de desserrement

Le raisonnement s'effectue à partir de particules sphériques.

Dans la zone des petits grains dominants (tableau 1), la théorie liée à l'effet de *paroi* considère que la perturbation générée par la paroi d'une grosse particule de diamètre $d_1$ entourée de petites particules se traduit par un supplément de porosité localisé dans une couronne sphérique délimitée par deux sphères : la première a un diamètre $d_1$, la seconde a un diamètre $d_{hyp}$ (Eq. (7)) dont l'expression fait intervenir un paramètre $k_{P2}$ déterminé (Eq. (8)) de façon à satisfaire la condition aux limites suivante : $b_{TPDM}(1) = 1$. La compacité des petites particules dans la zone perturbée par l'effet de paroi $\beta_2'(x)$ (Eq. (9)) et le coefficient d'effet de paroi $b_{TPDM}(x)$ (Eq. (10)) en sont ensuite déduits.

Tableau 1. Détermination du coefficient d'effet de *paroi* $b_{TPDM}$.

| | | | |
|---|---|---|---|
| Compacité virtuelle de la classe fine: $\beta_2$.<br>Indice des vides virtuel de la classe fine: $e_2^v$. | $\alpha_2$: compacité réelle de la classe fine.<br>$e_2$: indice des vides réel de la classe fine.<br>$K$: indice de compaction. | $\beta_2 = \alpha_2 \dfrac{(1+K)}{K}$<br>$e_2^v = (1+e_2)\dfrac{K}{(1+K)} - 1$ | (6) |
| Diamètre de la sphère délimitant la zone perturbée par l'effet de paroi autour d'une grosse sphère: $d_{hyp}$. | 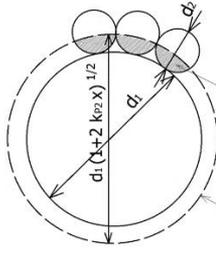 | $d_{hyp} = d_1\sqrt{1 + 2\,k_{P2}\,x}$ | (7) |
| Coefficient $k_{P2}$ calculé pour respecter la condition aux limites $b_{TPDM}(1) = 1$. | | $(\beta_2 - 6)(1 + 2k_{P2})^{\frac{3}{2}} + \dfrac{9}{2}k_{P2}^2 + 18\,k_{P2} + 5 = 0$ | (8) |
| Compacité des petites particules dans la zone perturbée par l'effet de paroi: $\beta'_2$. | $\beta'_2(x) = \dfrac{\pi(1+x)}{4x\left((1+2k_{P2}x)^{\frac{3}{2}}-1\right)\arcsin\left(\dfrac{x}{(1+x)}\right)}\left(2(1+2k_{P2}x)^{\frac{3}{2}}\right.$<br>$- 3(1+2k_{P2}x)\left(1 + \dfrac{k_{P2}x}{(1+x)}\right) + \left(1 + \dfrac{k_{P2}x}{(1+x)}\right)^3 - x^3$<br>$\left.+ \dfrac{3k_{P2}x^3}{(1+x)} + x^3\left(1 - \dfrac{k_{P2}}{(1+x)}\right)^3\right)$ | | (9) |
| Coefficient d'effet de paroi: $b_{TPDM}$. | $b_{TPDM}(x) = \dfrac{\left(\beta_2 - \beta'_2(x)\right)}{(1-\beta_2)}\left[(1+2k_{P2}x)^{\frac{3}{2}} - 1\right]$ | | (10) |

Tableau 2. Détermination du coefficient d'effet de *desserrement* $a_{TPDM}$.

| | | | |
|---|---|---|---|
| Compacité virtuelle de la classe grossière: $\beta_1$.<br>Indice des vides virtuel de la classe grossière: $e_1^v$. | $\alpha_1$: compacité réelle de la classe grossière.<br>$e_1$: indice des vides réel de la classe grossière. | $\beta_1 = \alpha_1 \dfrac{(1+K)}{K}$<br>$e_1^v = (1+e_1)\dfrac{K}{(1+K)} - 1$ | (11) |
| Diamètre de la cellule de référence sphérique pour étudier l'effet de desserrement: $d_{hyp}$. | 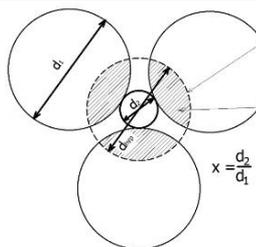 | $d_{hyp} = d_2\sqrt{1 + \dfrac{2\,k_{P1}}{x}}$ | (12) |
| | | $(\beta_1 - 6)(1 + 2k_{P1})^{\frac{3}{2}} + \dfrac{9}{2}k_{P1}^2 + 18\,k_{P1} + 5 = 0$ | (13) |
| Compacité des particules grossières dans la cellule de référence sphérique: $\beta''_1$. | $\beta''_1(x) = \dfrac{\pi(1+x)}{4\left(1+\dfrac{2k_{P1}}{x}\right)^{\frac{3}{2}}\arcsin\left(\dfrac{1}{(1+x)}\right)}\left(2\left(1 + \dfrac{2k_{P1}}{x}\right)^{\frac{3}{2}}\right.$<br>$- 3\left(1+\dfrac{2k_{P1}}{x}\right)\left(1 + \dfrac{k_{P1}}{(1+x)}\right) + \left(1 + \dfrac{k_{P1}}{(1+x)}\right)^3 - \dfrac{1}{x^3}$<br>$\left.+ \dfrac{3k_{P1}}{x^2(1+x)} + \left(\dfrac{1}{x} - \dfrac{k_{P1}}{(1+x)}\right)^3\right)$ | | (14) |
| Compacité des particules grossières dans la zone d'interférence plus globale: $\beta'_1$. | $\beta'_1(x) = \dfrac{\beta''_1(x)}{\beta''_1(x_0)\left\{1 + \dfrac{(x-x_0)}{(1-x_0)}\left(\sqrt[3]{\dfrac{2\beta''_1(1)}{\beta''_1(x_0)}} - 1\right)\right\}^3}\beta_1$<br>$x_0$: rapport critique de taille de caverne. | | (15) |
| Fraction volumique des petites particules dans un volume unité à l'eutectique: $\phi_2^*$. | $\phi_2^*(x) = \beta_2 + \left((1-\beta_2)(1-b_{TPDM}(x))-1\right)\beta'_1(x)$<br>Eutectique : point de passage dans la zone des « grains fins dominants ». | | (16) |
| Coefficient d'effet de desserrement: $a_{TPDM}$. | $a_{TPDM}(x) = \dfrac{(\beta_1 - \beta'_1(x))}{\phi_2^*(x)}\; if\; x \geq x_0\; et\; a_{TPDM}(x) = 0\; if\; x \leq x_0$ | | (17) |

Dans la zone des grosses particules dominantes (tableau 2), la théorie liée à l'effet de *desserrement* fait intervenir un rapport critique de tailles de caverne $x_0$. Si $x \leq x_0$, une fine particule peut s'insérer entre des gros grains sans modifier la structure de leur empilement. Si $x > x_0$, leur structure se décompacte localement. Pour assurer la continuité avec l'effet de *paroi* dans le cas où $x = 1$ (une seule taille de particules), une cellule de référence sphérique est utilisée pour calculer la compacité des grosses particules affectées par l'effet de desserrement : $\beta_1''(x)$ (Eq. (14)). Un calcul plus global est ensuite mené dans une zone dite d'interférence. Il permet d'aboutir à une compacité $\beta_1'(x)$ (Eq. (15)). Si $x = x_0$, $\beta_1'(x_0)/\beta_1 = 1$ : les petites particules n'impactent pas le squelette des grosses. Si $x = 1$, $\beta_1'(1)/\beta_1 = 0.5$ : les fractions volumiques des deux classes granulaires sont égales pour des particules identiques. Un raisonnement est ensuite effectué à l'eutectique (Eq. (16)) permettant un passage sans discontinuité de la zone des gros grains dominants vers celle des grains fins dominants. Le terme eutectique est utilisé en référence aux eutectiques moléculaires pour lesquels les petites molécules viennent se loger dans le réseau pseudo-cristallin des grosses molécules. Le coefficient d'effet de desserrement est ensuite déterminé (Eq. (17)).

$x_0$ évolue généralement entre 0 pour des particules très rugueuses aux formes très irrégulières et 0,2 pour des sphères sans friction (Roquier, 2024). Si ces dernières ne sont pas parfaitement lisses, $x_0$ est considéré égal à 0,17. Pour des particules anguleuses ou rugueuses, une valeur aux alentours de 0,03 est recommandée.

## 3. Le Modèle Théorique de Compacité (TPDM) pour les sols

Il convient maintenant de vérifier si les deux paramètres $K$ et $x_0$ sont adaptés à l'étude de l'indice des vides minimum de mélanges sable-fines non plastiques.

### 3.1. Le concept du rapport critique de taille de caverne $x_0$ pour les sols

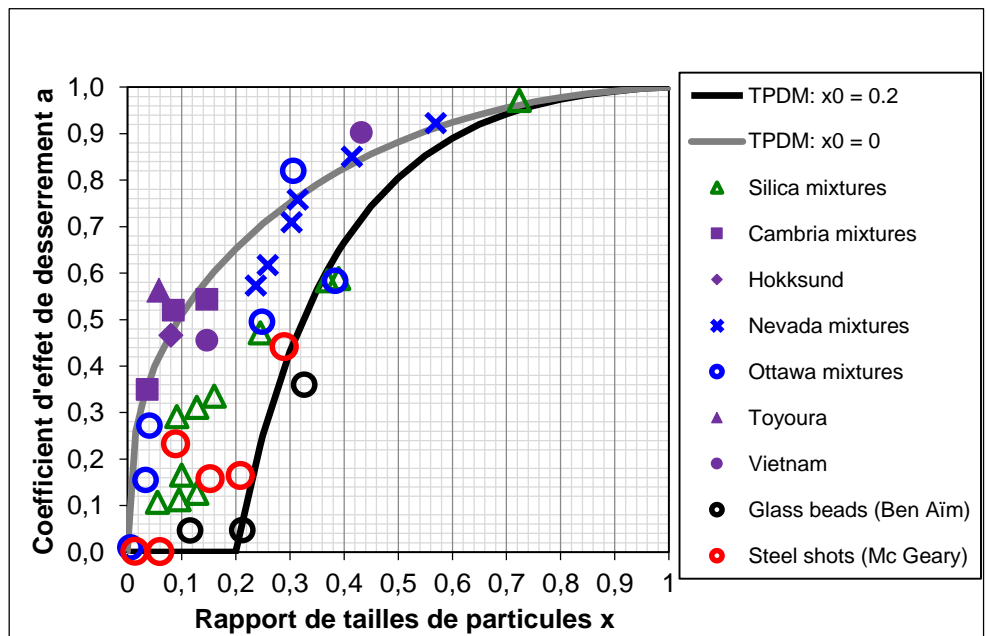

Figure 1. Coefficient d'effet de *desserrement vs* rapport de tailles de particules pour des sols granulaires typiques. Courbes du TPDM tracées avec $e_1$ = 0.56 et K = 9.

L'intervalle de $x_0$ étant compris entre 0 et 0,20, les points expérimentaux représentatifs du coefficient d'effet de desserrement $a$ en fonction du rapport de tailles $x$ sont censés être situés dans la zone délimitée par ces deux bornes. Le rapport de tailles pour les sols est défini comme étant égal à $x = d_{50}/D_{50}$ où $d_{50}$ et $D_{50}$ sont respectivement les tailles de particules moyennes pour les fines seules et pour le sable seul. Les données collectées dans (Chang *et al.*, 2015) sont relatives à des sols typiques dans le domaine des mélanges granulaires binaires : mélanges Cambria, mélanges Silica, Hokksund, mélanges Nevada, mélanges Ottawa, Toyoura, Vietnam. Deux jeux de données concernant des particules sphériques ont également été rajoutés à titre de comparaison : le premier relatif aux billes de verre (Ben Aïm, 1970), le second aux grenailles d'acier (McGeary, 1961) (Fig. 1). Les différents types de sols génèrent une dispersion assez forte des points. Ces derniers sont effectivement situés dans l'intervalle défini par $x_0 = 0$ et $x_0 = 0.2$ ce qui démontre qu'une modélisation basée sur ce paramètre est pertinente.

### *3.2. L'indice de compaction appliqué aux sols*

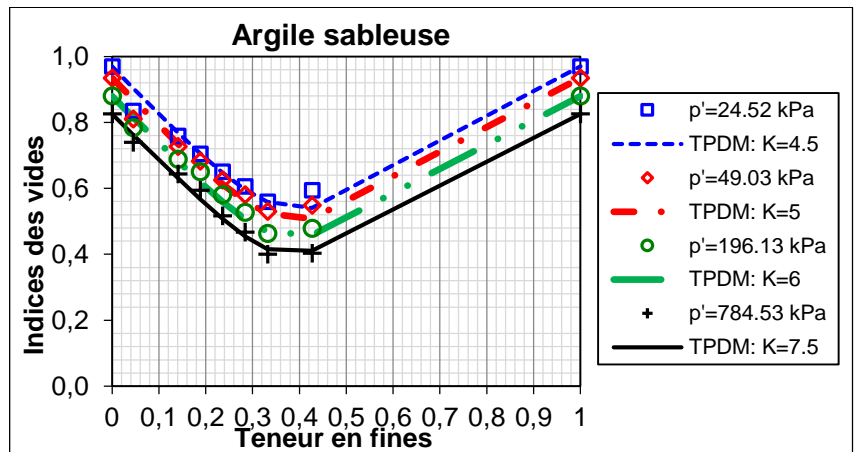

Figure 2. Comparaison entre les indices de vides mesurés sur une argile sableuse en fonction de la teneur en fines pour différentes pressions oedométriques (données de (Monkul et Ozden, 2007)) et estimations du TPDM pour différents indices de compaction *K*.

Monkul et Ozden (Monkul et Ozden, 2007) ont produit une étude très complète concernant l'évolution de l'indice des vides d'une matrice composite constituée de sable séché et de kaolinite en fonction d'une pression oedométrique évoluant entre 24,52 kPa et 784,53 kPa. Le pourcentage de fines $f_c$ varie entre 4,5% et 42,7% alors que l'indice des vides expérimental minimum est obtenu pour $f_c = 33.20\%$ (Fig. 2). Le rapport de tailles est très contrasté puisque $x = 0.01$. Le TPDM fournit des estimations s'accordant bien avec les données expérimentales pour un rapport critique de taille de caverne $x_0 = 0$. L'indice de compaction $K$ est donc en mesure de restituer l'effet généré par la pression oedométrique sur l'indice des vides du mélange testé, permettant d'envisager son utilisation pour estimer la variation de volume en cas de compactage d'un sol d'un état lâche vers un état dense.

### 4. Évaluation du Modèle Théorique de Compacité pour les sols : résultats et discussions

Le TPDM va être évalué sur la base de trois séries de mélanges dénommés (i) mélanges « Cambria – Nevada » ou « Nevada » (Lade *et al.*, 1998), (ii) mélanges « Quartz sand » (Liu *et al.*, 2020), (iii) mélanges « Sand – Fly ash » (Campbell *et al.*, 1983). Les premiers sont des combinaisons de sol de Cambria à particules rondes et de sol de Nevada à particules fines anguleuses. Le deuxième est un sable de quartz anguleux, tamisé pour offrir plusieurs classes granulaires mélangées entre elles. Enfin, le troisième est constitué de manière à améliorer les propriétés physiques d'un sol à texture grossière par l'ajout de cendres volantes, permettant ainsi de combiner des particules sphériques. Une étude menée en 2022 (Roquier, 2022) sur 26 mélanges binaires de sols a montré qu'un indice de compaction $K = 4.5$ pouvait être choisi en première estimation : il s'agit de la valeur retenue. Par ailleurs, les valeurs proposées dans le paragraphe 2.3. pour le rapport critique de taille de caverne $x_0$ en fonction de la forme et de l'état de surface des particules conduisent aux choix suivants : $x_0 = 0.03$ pour les mélanges (i) et (ii), $x_0 = 0.17$ pour le mélange (iii).

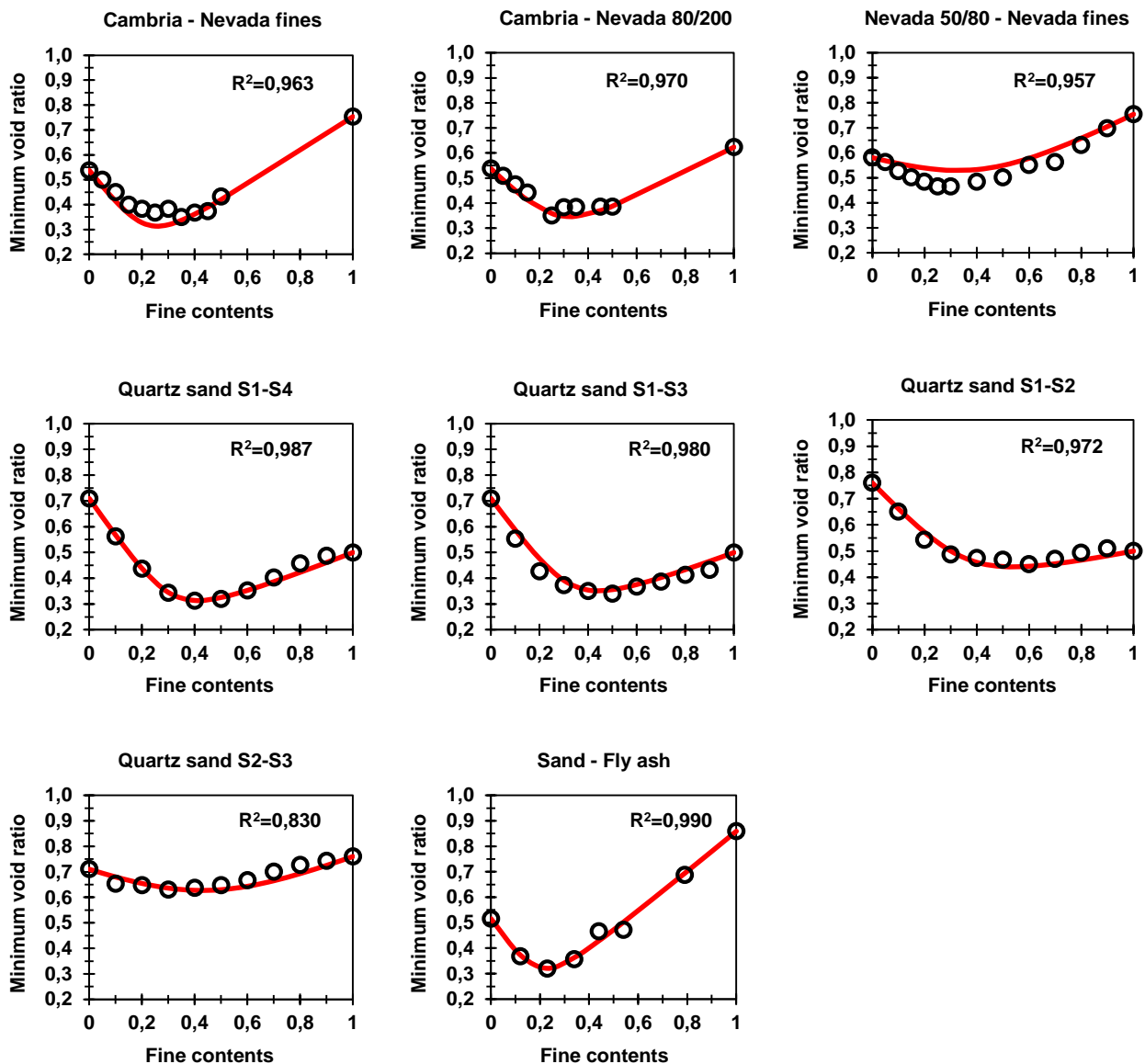

Figure 3. Comparaison des indices des vides minimum mesurés et estimés.

Le coefficient de détermination est égal à 0,956 en moyenne (Fig. 3). Grâce à l'indice de compaction $K$, le TPDM est capable de capturer le comportement non linéaire de la variation de l'indice des vides du mélange binaire exprimé en fonction de la fraction volumique des particules fines. Le choix du rapport critique de taille de caverne $x_0$ dépend du type de particules Grosses/Fines mélangées. $x_0$ vaut 0.17 pour les mélanges R/R (Rondes/Rondes) et 0.03 pour les mélanges A/A (Anguleuses/Anguleuses) et R/A. La valeur de 0.17 est très proche de celle mentionnée par Mc Geary (McGeary, 1961) dans son ouvrage de référence consacré à l'empilement des particules sphériques, qui vaut 0.15.

## 5. Conclusions

Avec le TPDM, on dispose d'un outil théorique assez avancé pour l'étude des mélanges granulaires, que ce soit dans les domaines des bétons, des asphaltes, des micro-poudres, des empilements cristallins et maintenant des sols.

Le TPDM constitue un outil générique et précis, applicable pour tous les rapports de tailles et pour des formes et des états de surface de particules variés. Ce modèle de forme fermée, c'est-à-dire dont les paramètres sont fixés à l'avance et adaptés à des empilements très diversifiés, est entièrement caractérisé par les proportions volumiques et les compacités ou indices des vides des classes granulaires fines et grossières, le rapport de tailles $x$, l'indice de compaction $K$ représentatif de l'état de densification de l'empilement (serré ou en vrac par exemple), et le rapport critique de taille de caverne $x_0$ à partir duquel se produit l'effet de desserrement et qui dépend de l'état de surface et de la forme des particules. Ainsi, $x_0$ vaut 0.17 pour les mélanges R/R (particules Rondes/Rondes) et 0.03 pour les mélanges A/A (Anguleuses/Anguleuses) et R/A. Validé ici pour estimer l'indice des vides minimum de mélanges sable-fines, des études complémentaires devront permettre de vérifier son application dans le cas d'un indice des vides maximum.

Il est à noter que le modèle est basé uniquement sur des principes physiques (eutectique par exemple) et qu'il ne subsiste plus de paramètres ajustables depuis que $K$ et $x_0$ ont été calibrés au fil des études (Roquier, 2022 et 2024). Ces derniers sont en particulier entièrement définis physiquement.

## 6. Références bibliographiques